# Complex structures of different $CaFe_2As_2$ samples

Bayrammurad Saparov*,[1], Claudia Cantoni[1], Minghu Pan[1], Thomas C. Hogan[2], William Ratcliff II[3], Stephen D. Wilson[2], Katharina Fritsch[4], Bruce D. Gaulin[4,5,6] & Athena S. Sefat[1]

[1]Materials Science and Technology Division, Oak Ridge National Laboratory, Oak Ridge, TN 37831, USA

[2]Department of Physics, Boston College, Chestnut Hill, MA 02467, USA

[3]NIST Center for Neutron Research, Gaithersburg, MD 20899-6102, USA

[4]Department of Physics and Astronomy, McMaster University, Hamilton, Ontario L8S 4M1, Canada

[5]Brockhouse Institute for Materials Research, McMaster University, Hamilton, Ontario L8S 4M1, Canada

[6]Canadian Institute for Advanced Research, 180 Dundas St W, Toronto, Ontario M5G 1Z8, Canada

*Correspondence to saparovbi@ornl.gov

**Abstract:** The interplay between magnetism and crystal structures in three $CaFe_2As_2$ samples is studied. For the nonmagnetic quenched crystals, different crystalline domains with varying lattice parameters are found, and three phases (orthorhombic, tetragonal, and collapsed tetragonal) coexist between $T_S$ = 95 K and 45 K. Annealing of the quenched crystals at 350°C leads to a strain relief through a large (~1.3 %) expansion of the *c*-parameter and a small (~0.2 %) contraction of the *a*-parameter, and to local ~0.2 Å displacements at the atomic-level. This annealing procedure results in the most homogeneous crystals for which the antiferromagnetic and orthorhombic phase transitions occur at $T_N/T_S$ = 168(1) K. In the 700°C-annealed crystal, an intermediate strain regime takes place, with tetragonal and orthorhombic structural phases coexisting between 80 to 120 K. The origin of such strong shifts in the transition temperatures are tied to structural parameters. Importantly, with annealing, an increase in the Fe-As length leads to more localized Fe electrons and higher local magnetic moments on Fe ions. Synergistic contribution of other structural parameters, including a decrease in the Fe-Fe distance, and a dramatic increase of the *c*-parameter, which enhances the Fermi surface nesting in $CaFe_2As_2$, are also discussed.

## Introduction

High-temperature superconductivity (HTS) in many of the cuprates and all of the iron-based superconductors (FeSC) is found in a narrow chemical composition region near an antiferromagnetic 'parent' compound. Upon electron- or hole-doping of the parent, the antiferromagnetic transition temperature ($T_N$) decreases, then a superconducting dome emerges. In fact, the discovery of HTS in cuprates was made in ~ 8% barium-substituted $La_2CuO_{4+\delta}$ ($T_N \approx$ 275 K) resulting in a superconducting transition temperature of $T_C$ = 40 K[1], and in FeSC in ~ 10% fluorine-doped LaFeAsO ($T_N \approx$ 145 K) producing $T_C$ = 26 K[2]. Thus, fundamental studies of the parents may give clues about the puzzling origin of HTS. Our study here focuses on the $CaFe_2As_2$ ('122'), the magnetic and structural properties of which have been a subject of interest since the discovery of FeSCs[2-4]. The superconductivity in 122 parents, which adopt the tetragonal $ThCr_2Si_2$-type structure at room-temperature, can be induced through the suppression of $T_N$ ($T_S$) by applying pressure[5] or by chemical substitution on the alkaline-earth metal[3] or Fe site[4]. For $CaFe_2As_2$, a range of $T_N$ ($T_S$) values can be found in literature that is dependent on the sample's preparation or post-growth treatment. For example, crystals grown out of FeAs self-flux are reported to give $T_N$ = 165 K[6] or no $T_N$[7], dependent on quench-temperature. Crystals grown out of Sn flux have $T_N \approx$ 170 K[8-10] (with small percentage Sn incorporation), while a polycrystalline sample shows $T_N$ = 160 K[11]. In addition, thermal-annealing of $CaFe_2As_2$ crystals is reported to cause significant changes in bulk properties, with a shift up to ~ 50 K in $T_N$[7], even though 122 composition and room-temperature $ThCr_2Si_2$ structure remain unchanged. For the annealed crystals, the dramatic changes were argued to be caused by the formation of nanoscale $Fe_2As$-rich precipitates that cause internal pressure on 122 phase[7].

There is a rich temperature (T) – pressure (P) phase diagram for $CaFe_2As_2$ (see Figure S1), in which various phases, in addition to being accessible under applied pressure, can be also reached by annealing[7]. This work focuses on three different $CaFe_2As_2$ samples in this phase diagram: as-grown 'p1' sample, obtained by air-quenching $CaFe_2As_2$ at 960°C, 'p2', which results from annealing of the as-grown crystals at 700 °C, and 'p3', acquired by annealing of the as-grown crystals at 350 °C. We report on the details of nuclear and magnetic structures, and local structural distortions that give rise to the dramatic changes in magneto-elastic properties. We find that the magnetic behavior of different $CaFe_2As_2$ samples is determined by several structural parameters that are of crucial importance, including the Fe-As and Fe-Fe distances, anionic As height from Fe plane, and the *c*-lattice parameters.

## Results

**Physical properties**. Temperature-dependence of bulk physical properties is shown in Figure 1. Transition temperatures (denoted by $T^*$), which were determined from $d\rho/dT$, $C(T)$ and Fisher's $d(\chi T)/dT$, are comparable; see the summary in Table 1. From $\rho(T)$ measured upon warming and cooling, the most prominent hysteresis is seen within p1 for which $\Delta T^*$ is ~ 5 K. The largest change in $T^*$ is observed between p1 ($T^* \approx 95$ K) and p3 ($T^* \approx 170$ K). Sample P2 ($T^* \approx 122$ K) gives a $T^*$ value that lies in between that for p1 and p3; these results corroborate with structural parameter changes discussed below. The residual resistivity ratio (RRR = $\rho_{300 K}/\rho_{2 K}$) values decrease with annealing: 8.0 for p1, 5.4 for p3, and 1.9 for p2. Assuming that crystal quality increases with annealing out defects, disorder, and other structural imperfections, one would instead expect an increase in RRR. Here, however, as the differences in $\rho(T)$ behavior and $T^*$ show, the annealing temperature has a great influence on the magnetic and structural properties of $CaFe_2As_2$.

The $T^*$ peaks in $C(T)$ data are a lot less intense and broad for p1 and p2, compared with the sharp peak for p3 (Figure 1c, and inset); the reasons for these features are in the crystal and magnetic structural transitions, which are presented below. The strong peak for p3 is due to the magnetic and structural transitions coinciding at $T^* = 168(1)$ K. For p2, the structural transition occurs over a very wide temperature range of 80 K to 120 K, hence the broad but noticeable feature in $C(T)$ at $T^* = 111(1)$ K. For p1, which does not undergo magnetic ordering, the X-ray diffraction data (see below) show that the structural collapse occurs in most of the domains in a 5 K temperature range. The Sommerfeld coefficients, from linear fits of $C/T$ vs $T^2$ below 10 K, are $\gamma \approx 14$, 8, and 6 mJ/(mol K$^2$) for samples p1, p2, and p3, respectively. The trends in $\gamma$ values are consistent with the trends in $T^*$, and indicate a decrease of low energy excitations with annealing. Although there are no reports on heat capacity for the FeAs grown $CaFe_2As_2$ crystals, for Sn-grown $CaFe_2As_2$, $\gamma = 8.2$ mJ/(mol K$^2$)[9].

**X-ray and Neutron diffraction**. Upon annealing, an increase in the room-temperature $c$-lattice parameters of $CaFe_2As_2$ crystals was noted in literature: as-grown crystals gave $c$ ~11.62 Å, while 400°C-annealed crystals showed $c$ ~ 11.8 Å[7,12]. Assuming that the as-grown p1 crystals are strained upon quenching from 960°C (flux-spin temperature), we would expect them to have an inhomogeneous strain distribution. Consistent with this expectation, we observe broad shoulders on the (0 0 $l$) Bragg peaks collected for single crystals (see Figure 2a for (0 0 8) peak). The fit of the strongest peaks at room temperature gives $c = 11.574(1)$ Å, whereas the estimated fitting of the shoulders gives $c = 11.688(9)$ Å. The fact that the (0 0 $l$) peaks are broadened, but not split, suggests that sample p1 has crystalline domains with $c$-lattice parameters varying between ~ 11.574 and 11.688 Å. This finding highlights the sensitivity of the $CaFe_2As_2$ crystal structure, showing that there are many energy minima in the energy landscape of this compound. Upon cooling, the Bragg reflections broaden further, and a second shoulder appears at 85 K. At this temperature, we observe emergence of a (0 0 8)$_{CT}$ peak for the collapsed-tetragonal (CT) phase. The structural collapse in sample p1 is manifested as a dramatic decrease in the $c$-parameter accompanied by an increase in the in-plane lattice parameter (Figures 3 and S2). Interestingly, we observe that a small fraction (< 3 % based on $I_{(008)T}/I_{(008)CT}$) of the shoulder is still present down to 45 K. Upon cooling p1, a small peak is also evident at ~ 150 K that corresponds to the orthorhombic (O) phase with larger $c \approx 11.64$ Å (Figure 2a). In addition to the slightly increased $c$-lattice parameter, the reported characteristic feature of the (0 0 8)$_O$ peak is that it shows no thermal contraction[13] (see also neutron diffraction results below), and indeed, we find that the (0 0 8)$_O$ position remains approximately unchanged upon cooling. The relative fraction of the O phase, estimated from the intensities of the (0 0 8) peaks for the CT and O phases ($I_{(008)O}/I_{(008)CT}$), is less than 3 %. The heat capacity data do not show any detectable feature at ~ 150 K that can be attributed to the formation of this phase, because the fraction of the crystal that undergoes the O transition is small. In conclusion, for p1, the O, T, and CT phases coexist within the various crystalline domains, in the temperature range of 45 to 90 K.

Annealing of sample p1 is expected to relieve strain and lead to more homogeneous, chemically-ordered, and strain-free crystals. Indeed, sample p2 gives sharp (0 0 $l$) Bragg reflections at room temperature (Figure 2b), with no accompanying shoulder or impurity phases. There is a significant increase in the $c$-parameter to 11.6665(7) Å at room temperature, ~ 0.8 % increase compared to the majority of domains in p1. Upon cooling sample p2, a large temperature window of coexistence of T and O phases is observed. The gradual structural transition occurs over 40 K, upon cooling from 120 K to 80 K. This unusual behavior can explain the very small peak at 111(1) K noted in heat capacity. Such a wide transition is indicative of strain that is not completely relieved. X-ray diffraction shows that p3 crystals are the most ordered (Figure 2c). The strain is relieved with an accompanied expansion of the $c$-lattice parameter to 11.7262(8) Å at room temperature, which corresponds to ~ 1.3 % increase compared to sample p1; this value is also 0.5 % larger than the $c$-parameter for sample p2. For p3, there is a sharp transition to the O phase at 170 K, without any detectable coexistence region. Additionally, the structural transition is found to be at the same temperature as the magnetic transition (see Figure S2), hence the strong peak in the heat capacity. Trends in the lattice parameters, upon cooling and warming, are displayed in Figure 3 and Figure S2. For clarity, the $c$-parameter of the O phase (<3 %) in p1 is not shown. The maximum hysteresis amounts for the transition temperatures for p1 and p2 are ~ 40 K and 15 K, respectively. For p3, above the structural transition temperatures, the temperature dependence of $c$-parameter is linear, whereas p2 and p3 show curvatures; such trends may also be related to strain. Another interesting observation is that the structural collapse in p1 occurs after the $c$-parameter drops below 11.2 Å, and because the less-strained crystals have higher $c$-lattice parameters, the formation of CT phase is not triggered in them.

The X-ray data described above is corroborated by the neutron diffraction experiments (Figure S2). The structural collapse in p1, which is a first order transition, shows a region of two-phase coexistence; however, at the lowest temperature measured (4 K) only a small (<1%) volume fraction of the sample remained in the uncollapsed tetragonal phase. This remnant volume fraction is difficult to determine precisely due to strain-induced broadening of the mosaic of the uncollapsed domains within the larger collapsed ma-trix. Consistent with the known quenching of the antiferromagnetic iron moment in the CT phase, Figure S2a does not reveal long-range antiferromagnetic order at the known superlattice reflection $Q = (1\ 0\ 3)$ in sample p1. We note that very weak magnetic scattering was apparent in our sample (data not shown) suggestive of an O distortion in a small volume fraction. The fact that we observe small fractions of uncollapsed T and O phases along with the dominant CT phase are in complete agreement with the above X-ray diffraction data, which estimate the relative content of each of these phases below 3 %. For p3, the structural and magnetic phase behavior reverts to the nominal $T_S$ and $T_N$ observed for the Sn-flux grown CaFe$_2$As$_2$ crystals. Below $T_S = 168(1)$ K, a first order distortion for T – O transition occurs that is accompanied by the onset of long-range antiferromagnetic order. The intensity of the (1 0 3) magnetic peak is proportional to the magnetic order parameter squared ($M^2$) and is plotted in Figure S2b. The sharp, first-order onset of magnetic order in this system is consistent with the strong magneto-elastic coupling expected below the tricritical point.

The results of Rietveld fits of the PXRD data are summarized in Tables 2 and 3. A representative powder X-ray pattern and structure refinement of p1 is shown in Figure S3. Due to the extreme pressure sensitivity of CaFe$_2$As$_2$, grinding crystals into powders led to shifts of peak positions and changes in peak shapes compared to the data obtained for single crystals. The observed difference in the $c$-parameters from single crystals and powders, hence, is attributed to the applied pressure when grinding, and the fast cooling and heating, which samples experience when immersed in liquid nitrogen. Confirming the expected behavior, unit cell volume increases with annealing, with final cell volume difference between p1 and p3 measuring up to ~ 2 Å$^3$ at room temperature.

It is important to note that the three samples studied here represent the states of CaFe$_2$As$_2$ below 1.7 GPa at 300 K, the critical pressure at which the transition to CT phase occurs. Recent theoretical work[14] has highlighted the crucial importance of several structural parameters, first of all, the iron-pnictogen bond length, which has an important role in determining localized (longer Fe-As length) or itinerant (shorter Fe-As length) nature of iron electrons. In addition, it has been proposed that the Fe-As bond length determines the chemical pressure on Fe, with higher pressures leading to a loss of the local moment, and therefore, it is a gauge of the magnitude of the local magnetic moment[15,16]. In the CaFe$_2$As$_2$ samples, we see an increase in $d_{Fe-As}$ with annealing from $d_{Fe-As} = $ 2.373(1) Å in p1 to $d_{Fe-As} = $ 2.3826(6) Å in p3, which implies a more localized nature of Fe electrons, and higher local magnetic moments on Fe ions for p3. Therefore, the observed magnetic behavior in CaFe$_2$As$_2$ samples, namely, increasing $T_N$ with annealing is directly correlated with the rising local Fe moment due to longer Fe-As bond length. We observe a slight decrease of the Fe-Fe distance, which is also important in determining the behavior of iron-based materials[14,15], from $d_{Fe-Fe} = $ 2.7567(1) Å in p1 to $d_{Fe-Fe} = $ 2.7513(1) Å in p3, consistent with increasing $T_N$. The third parameter is the tetrahedral angle As-Fe-As, which controls the crystal-field effect and the orbital occupancies[14,17]. We find that the As-Fe-As bond angles in p1 CaFe$_2$As$_2$ crystal are 110.50(5)° and 108.96(5)°, with annealing, the FeAs$_4$ tetrahedra become less distorted, leading to almost regular tetrahedral angles of 109.46(2)° and 109.48(2)° in p3. Although it is known that the distortion of the FeAs$_4$ tetrahedron enhances the crystal-field splitting, a direct connection between this parameter and the antiferromagnetic ordering temperature has not been found so far; however, it has been reported that the crystal-field splitting is more enhanced in the pressurized, nonmagnetic CaFe$_2$As$_2$ phase[17]. The $c$-axis, another parameter, is a gauge for pressure in CaFe$_2$As$_2$, and pressure has a dramatic influence on the Fermi surface. Reducing the $c$-parameter leads to a gradual shrinkage of the inner hole and electron cylinders, and a consequent reduction of the nesting features, resulting in a decrease of the magnetization[17]. Consistent with this, p1, the most pressurized sample with the smallest $c$-parameter, likely does not possess nesting features in the Fermi surface, and therefore, transitions into a nonmagnetic CT state at low temperatures. The large expansion of $c$-parameter in p3 leads to a ~ 0.04 Å increase in the interlayer distance measured by the distance between two As atoms in adjacent layers (Table 3), which suggests increased 2D character. Another structural parameter with great influence on transition temperatures is the anion arsenic height ($h_{As}$)[18], which also follows the trend in $T_N$ in the three samples of CaFe$_2$As$_2$ (Table 3).

**Microscopy**. Results of scanning tunneling microscopy (STM) and scanning tunneling spectroscopy (STS) studies of CaFe$_2$As$_2$ are presented in Figure S4 (see Supporting Information for detailed discussions). STM/S data indicate that there is a strong inhomogeneity of bosonic excitations in p1, which suggests a weak or non-antiferromagnetic ordering. Consistent with the X-ray and neutron diffraction results, there is a strongly inhomogeneous strain distribution in p1 with the minority of domains remaining in the uncollapsed magnetic state at low temperatures, while the majority of crystalline domains transition into the collapsed nonmagnetic state. In contrast, obtained data for p3 suggest presence of strong bosonic excitations, which originate from spin excitations in the antiferremagnetic state.

The scanning transmission electron microscopy (STEM) results at room temperature, on the local atomic structure of different samples of CaFe$_2$As$_2$, are presented in Figures 4 and 5. For sample p1, the atomic resolution aberration-corrected STEM image using the high-angle annular dark-field detector (HAADF) mode, with the beam parallel to [1 0 0] and [0 0 1], respectively, are shown in Figure 4a and 4c. Here, the images' intensity, which is proportional to ~ $Z^2$, is a first indication of chemical homogeneity: Ca, Fe and As columns appear in dark, medium-light, and light shades of gray, respectively. Further indication of uniform chemical composition is given by the electron energy loss spectroscopy (EELS) elemental maps, shown in Figure 4b and 4d, acquired by scanning the beam over the square regions indicated in Figure 4a and 4c; similar results were obtained on samples p2 and p3 as well (not shown). In spite of the observed chemical uniformity, maps of the $a$-lattice parameter obtained from the images showed displacements of the atomic columns that were random and larger than the measurement uncertainty. Lattice spacing maps were obtained after atomic ($x_i$, $y_i$) coordinates were calculated using an iterative algorithm to locate the center of mass of each column in the HAADF image. An estimate of the local strain can be obtained by dividing the standard deviation for the lattice spacing distribution in one of the maps by the average lattice spacing. This gives a 2% value for p1 over an area of 50 × 40 areal unit cells, indicating a large local strain field. Upon annealing, the local strain is greatly reduced (below 1 %), but major changes are observed in the positions of atoms within the p2 and p3 structures. As shown in the HAADF images (Figure 5), in p2 and p3, the Fe sublattice is shifted with respect to the Ca and As sublattice along in-plane direction as compared with the p1 crystal. In the case of p2, a subset of the Ca and As sublattice is also shifted with respect to the rest of the crystal, in the same direction. The atomic displacements can occur along either the <1 0 0> or <0 1 0> direction giving rise to crystalline domains of 10 to 20 nm in size. The presence of nanoscale domains renders this type of lattice distortion very challenging for detection by bulk-sensitive X-ray and neutron diffraction techniques. Shifts of atomic planes in the annealed crystals lead to changes in the Fe-As bond lengths and angles. Simulation of these atomic displacements in a crystal structure software using the p1 structure from X-ray data as a reference gives comparable changes in the Fe-As bond length to those found from X-ray diffraction measurements: the average Fe-As bond lengths are $d_{Fe-As}$ = 2.375 Å in p2 and $d_{Fe-As}$ = 2.384 Å in p3. Similarly, the average values of bond angles point to more regular tetrahedra upon annealing, in agreement with the bulk diffraction results. At the local level, however, the FeAs$_4$ tetrahedra are most distorted in the case of sample p3, where due to the displacements of Fe atoms relative to As, all the resultant bond lengths differ from the original bond lengths in p1 crystal. For p2 crystal, the distance between As and Fe columns in the $x$ direction (see schematic drawings in Figure 5) alternate between small and large values in the plane labeled As1, Fe1 implying a displacement of the Fe atoms of ~ 0.2 Å, compared to less than 0.05 Å in p1. However, in the next layer (labeled As2, Fe2), there is no significant change in the distance between As and Fe columns. In p3, both As1-Fe1 and As2-Fe2 displacements occur out of phase, and have virtually an identical magnitude of ~ 0.1 Å. We note that in high-resolution aberration-corrected STEM, atomic-resolution images arise from a layer less than 10 nm in thickness, while the signal coming from the rest of the sample (typically ~30 nm thick) contributes to the background. Thus, our images probe a region of the sample most likely contained in a single domain (~30 nm in size) as opposed to one encompassing several domains; therefore, explaining the sensitivity to local displacements.

### Discussion

The large changes in the physical properties of CaFe$_2$As$_2$ crystals upon annealing are explained here through the details of bulk and local crystal structures. Quenching of the crystals from high temperature via centrifugation arrests the as-grown p1 crystals into a highly strained state where domains of varying $c$-lattice parameters in the range of 11.574 – 11.688 Å coexist at room temperature. In addition to a macroscopic component, this strain has also a microscopic component manifesting as static random displacements of the atoms from their equilibrium positions. As a consequence of the non-uniform strain distribution, there is evidence for strong inhomogeneity of bosonic excitations in STM/S, which originate from spin excitations. CaFe$_2$As$_2$ in sample p1 is disordered both structurally and electronically, and features a mixture of areas with weak bosonic excitations and no-bosonic excitations, preventing long-range antiferromagnetic order. The nonmagnetic nature of p1 is confirmed here by neutron diffraction results. Indeed, the highly inhomogeneous strain fields in p1 result in a tetragonal to nonmagnetic collapsed-tetragonal phase transition below 95 K, and in a temperature window between 45 K and 90 K, where all three CaFe$_2$As$_2$ known phases (tetragonal, collapsed tetragonal and orthorhombic) coexist. The annealed p2 undergoes a partial strain relief via collective, local atomic displacements at the microscopic level, and a significant ~ 0.8 % increase in the $c$-lattice parameter compared to p1. For p2, the distances between Ca/As and Fe columns along an in-plane direction show a small-large modulation in every second (100) atomic plane. The X-ray diffraction results indicate that the $c$-parameter distribution is uniform; however, upon cooling, p2 shows a 40 K coexistence region of tetragonal and orthorhombic phases from 120 K down to 80 K. Such a gradual transition and mixtures of phases in a large temperature range is a clear signature that although less strained, p2 is microscopically inhomogeneous. Annealing at the relatively low temperature to obtain p3 relieves strain most effectively, causing a macroscopic expansion of the $c$-parameter by 1.3 %, and a microscopic modulation of the distance between Ca/As and Fe columns in every (100) plane, corresponding to a rigid shift between Fe and Ca/As sublattices. P3 shows an abrupt transition from tetragonal to orthorhombic phase at 170 K denoted by a sharp peak in heat capacity. Spatial mapping of bosonic mode in p3 suggests that it is largely composed of regions with strong bosonic mode below 170 K, consistent with the long-range antiferromagnetic ordering found in neutron diffraction results.

The structural changes, most importantly the Fe-As bond length increase from $d_{Fe-As}$ = 2.373(1) Å in p1 to $d_{Fe-As}$ = 2.3826(6) Å in p3, suggest more localized electrons and increased local magnetic moments on iron ions for p3. The more localized Fe electrons and high local Fe magnetic moments, in turn, are responsible for the dramatic increase in the antiferromagnetic ordering temperature. Other structural parameters also contribute synergistically to this effect: annealing, Fe-Fe distances decrease from $d_{Fe-Fe}$ = 2.7567(1) Å in p1 to $d_{Fe-Fe}$ = 2.7513(1) Å in p3; the dramatic increase of the $c$-parameter with annealing enhances the Fermi surface nesting in p3 sample of CaFe$_2$As$_2$; the anion height $h_{As}$ increases with increasing $T_N$. Interesting to note that the observed superconductivity in pressurized CaFe$_2$As$_2$ is reported to be due to low temperature inhomogeneous phases[19]. Here, we also observe crystallographically multi-phase samples with inhomogeneous distribution of strain and bosonic excitations. Therefore, in addition to the structural parameters, the inhomogeneity itself may play an important role in the magnetism of these materials.

The gradual structural transition in sample p2 occurring over 40 K requires further investigations. Despite such a broad and gradual structural transition to the orthorhombic phase, the magnetization data for this crystal shows a much sharper transition at $T^*$ = 118(4) K. The fraction of the orthorhombic phase at 120 K on cooling and warming are less than 4.1 % based on ($I_{(008)O}/I_{(008)T}$) from X-ray diffraction results. Therefore, quite unusually, this result may indicate decoupling of the structural and magnetic transitions.

## Methods

**Synthesis and preparation.** CaFe$_2$As$_2$ crystals were synthesized by reacting pieces of Ca element with FeAs binary in a Ca:FeAs = 1:4 ratio. Reactions mixtures were heated to 1180 °C, homogenized for 24 h, and slowly cooled to 960 °C, at which point the flux was removed by centrifugation. These crystals are sample 'p1'. In order to reach other parts of the phase diagram (Figure S1), selected p1 crystals, with no visible surface flux, were sealed under vacuum in silica, and then annealed in pre-heated box furnaces at either 350 °C or 700 °C. After the annealing, each reaction tube was quenched in air. The 350 °C- and 700 °C-annealed crystals are called 'p3' and 'p2', respectively.

**Characterization**

*X-ray Diffraction.* A PANalytical X'Pert PRO MPD X-ray diffractometer, with monochromated Cu-K$\alpha$1 radiation, was used to carry out room- and low-temperature data collections. The (0 0 $l$) Bragg peaks of the flat-lying CaFe$_2$As$_2$ crystals at high angle regions were collected to study the structural evolution of different samples of CaFe$_2$As$_2$ upon cooling.

*Elemental analysis.* Elemental compositions of the samples were confirmed using the energy-dispersive X-ray spectroscopy (EDS) measurements performed on a Hitachi-TM3000 scanning electron microscope equipped with a Bruker Quantax 70 EDS system.

*Physical property measurements.* Magnetization data were collected using a Quantum Design Magnetic Property Measurement System (MPMS), in zero-field-cooled mode under 10 kOe and upon warming from 2 K in the *ab*-plane. The measurements of temperature-dependence of electrical resistivity, $\rho_{ab}(T)$, and heat capacity, $C(T)$, were carried out on a Physical Property Measurement System (PPMS). Commercial equipment referred to in this paper is identified for information purposes only, and does not imply recommendation of or endorsement by the National Institute of Standards and Technology, nor does it imply that the products so identified are necessarily the best available for the purpose.

*Neutron Diffraction.* Neutron diffraction measurements were performed on the BT-4 triple-axis spectrometer at the NIST Center for Neutron Research at the National Institute of Standards and Technology. Experiments were performed using a pyrolitic graphite (PG) monochromator and PG analyzer with a final energy $E_f$ = 14.7 meV. One PG filter was placed after the monochromator, one after the sample, and collimations were 40'-40'-40'-open before the monochromator, sample, analyzer, and detector, respectively. The sample was aligned within the ($h$ 0 $l$) scattering plane and loaded into a closed cycle refrigerator for measurements.

*Microscopy.* Scanning tunneling microscopy (STM) and scanning tunneling spectroscopy (STS) investigations were performed at 78 K for the in situ-cleaved CaFe$_2$As$_2$ crystals. The differential tunneling conductance spectra (dI/dV versus V) surveys were taken on a large surface area (200 nm × 200 nm), and for each crystal, averaged over 2500 curves.

For STEM investigation, crystals of CaFe$_2$As$_2$ were cleaved to obtain small slabs of 40 μm or less in thickness, subsequently glued to a Cu grid, and thinned to electron transparency (≈ 30 nm) by Ar ion milling at a voltage of 2 kV with a final cleaning step at 0.5 kV. As prepared TEM samples were heated up to 80 °C in high vacuum ($10^{-7}$ Torr) for several hours in order to remove hydrocarbon contamination from air exposure, and then rapidly loaded in the microscope. Oxygen contamination was monitored using EELS in the microscope. The microscope used for this study is a Nion UltraSTEM100 operating at 100 kV, equipped with a second generation 5$^{th}$ order probe aberration corrector, a cold emission electron gun, and a Gatan Enfina EEL spectrometer. Ion milling at voltages higher than 2.5 kV resulted in surface disorder consisting of amorphous clusters of Ca-O and Ca-F, which were discernible in HAADF images and EELS spectrum images.

## Acknowledgements

This work was supported by the Department of Energy, Basic Energy Sciences, Materials Sciences and Engineering Division. CC acknowledges support by ORNL's Shared Research Equipment User Program, which is sponsored by the Office of Science of Basic Energy Sciences, U.S. Department of Energy. MHP acknowledges the support of the Scientific User Facilities Division, Office of Basic Energy Sciences, U.S. Department of Energy, for the work conducted at the Center for Nanophase Materials Sciences in ORNL. SDW acknowledges support under NSF CAREER DMR-1056625. Work at McMaster University was supported by NSERC of Canada. The authors acknowledge and greatly appreciate discussions with Elbio R. Dagotto and Krzysztof Gofryk.

## Author contributions

B.S. prepared the samples, carried out PXRD, electrical resistivity, magnetic susceptibility and heat capacity measurements. C. C. performed STEM measurements. M. P. carried out STM and STS experiments. T. C. H., W. R. II and S. D. W. carried out neutron diffraction experiments. K. F. and B. D. G. contributed to X-ray diffraction experiments. A. S. S. initiated the project, assisted with the physical property measurements, supervised the experiments, analyzed the data and wrote the paper. All authors reviewed the manuscript. Everyone contributed to the data analysis, discussions, and writing of the manuscript.

## Additional information

**Supplementary information** accompanies this paper at http://www.nature.com/scientificreports
**Competing financial interests:** The authors declare no competing financial interests.

**Table 1** Transition temperatures for the different $CaFe_2As_2$ samples, determined from bulk properties.

| $CaFe_2As_2$ | $d\rho/dT$ on cooling | $d\rho/dT$ on warming | $d(\chi T)/dT$ on warming | $C(T)$ on cooling |
|---|---|---|---|---|
| p1 | 90(1) K | 95(1) K | 96(1) K | 92(1) K |
| p2 | 118(2) K | 122(1) K | 118(4) K | 111(1) K |
| p3 | 168(1) K | 170(1) K | 171(1) K | 168(2) K |

**Table 2** For the three samples of $CaFe_2As_2$, room-temperature lattice parameters ($a$, $c$), cell volumes ($V$), $z$ coordinates for As ($z_{As}$), arsenic heights from Fe layer ($h_{As}$) and goodness-of-fit (GoF) values.

| Sample | $a$, Å | $c$, Å | $V$, Å$^3$ | $z_{As}$ | $h_{As}$, Å | GoF |
|---|---|---|---|---|---|---|
| p1 | 3.8986(2) | 11.598(1) | 176.28(2) | 0.3666(2) | 1.346 | 1.13 |
| p2 | 3.8925(2) | 11.680(1) | 176.97(2) | 0.3663(2) | 1.363 | 2.44 |
| p3 | 3.8909(1) | 11.7682(7) | 178.16(1) | 0.36691(9) | 1.380 | 1.82 |

**Table 3** Selected interatomic distances (Å) and angles (°) in the three samples of $CaFe_2As_2$.

| | Fe–As×4 (Å) | Fe–Fe×4 (Å) | As–As (Å) |
|---|---|---|---|
| p1 | 2.373(1) | 2.7567(1) | 3.094(3) |
| p2 | 2.373(1) | 2.7524(1) | 3.123(3) |
| p3 | 2.3828(6) | 2.7513(1) | 3.133(3) |

| | As–Fe–As (°) | Fe–As–Fe (°) |
|---|---|---|
| p1 | 108.96(5), 110.50(5) | 71.04(4), 110.50(5) |
| p2 | 109.12(5), 110.17(5) | 70.88(4), 110.17(5) |
| p3 | 109.46(2), 109.48(2) | 70.53(2), 109.49(2) |

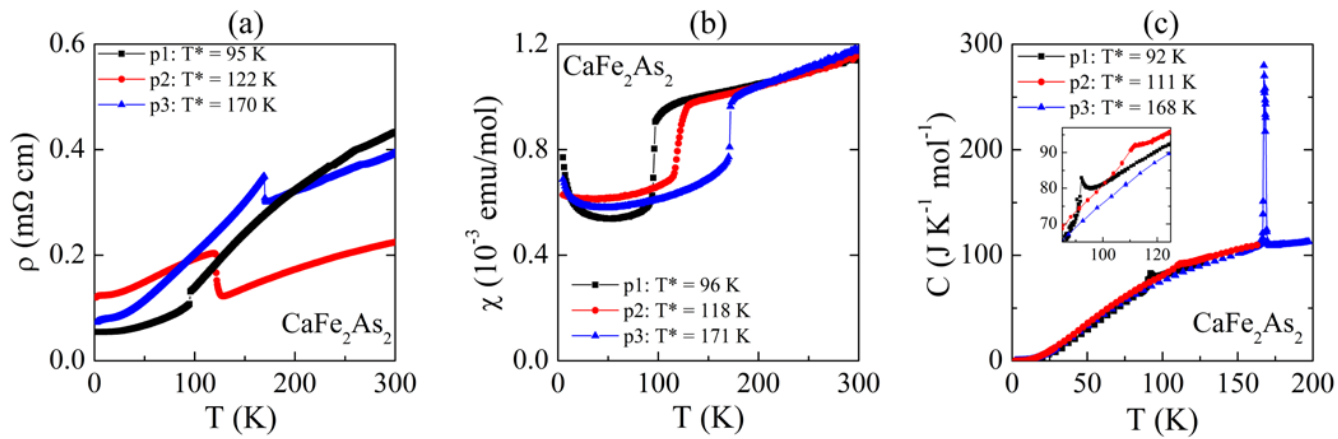

**Figure 1**. Temperature dependence of (a) resistivity, (b) magnetization and (c) heat capacity for the three samples of $CaFe_2As_2$.

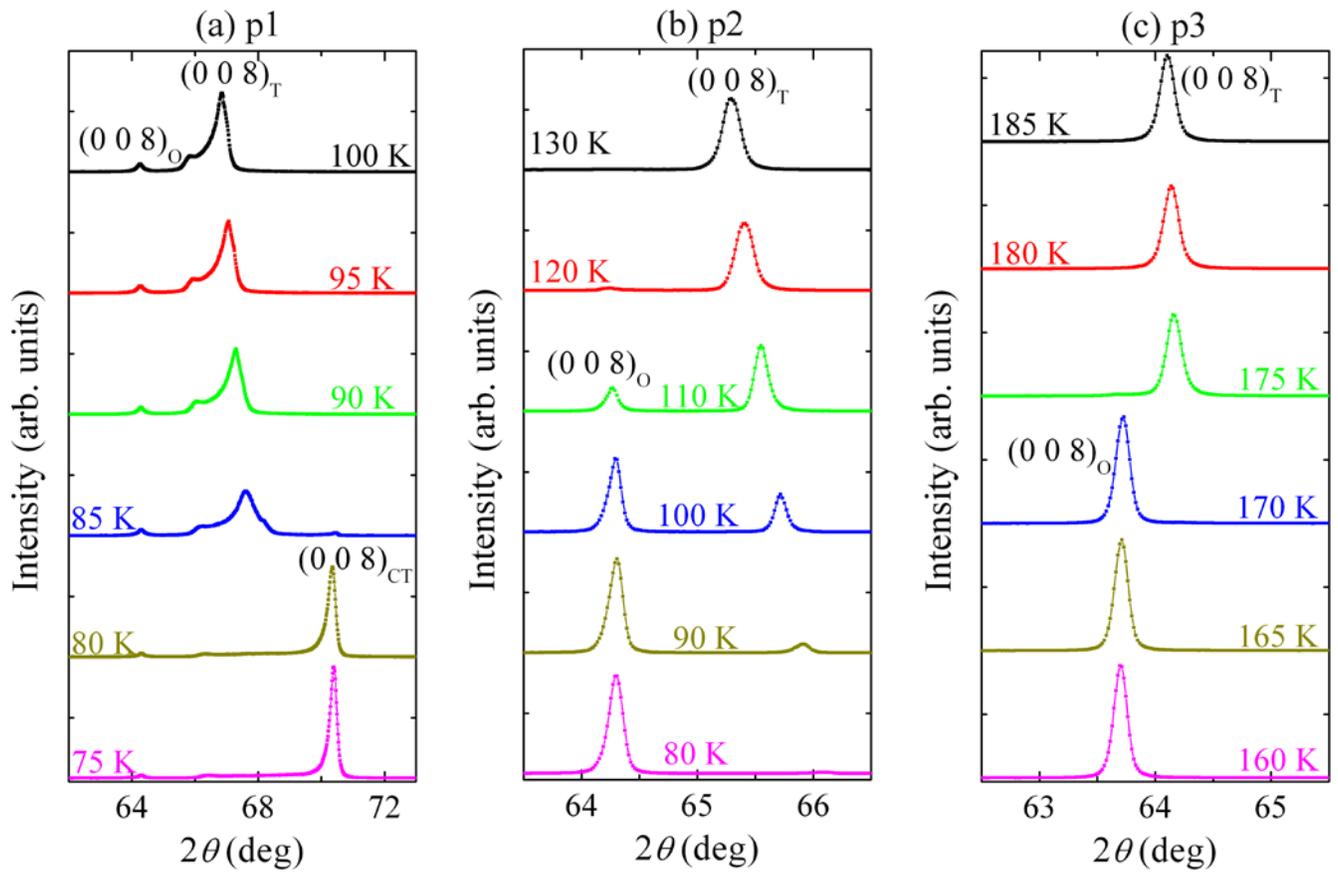

**Figure 2**. Temperature dependence of (0 0 8) peaks near the transition temperatures in (a) p1, (b) p2 and (c) p3 samples of $CaFe_2As_2$. The as-grown p1 crystal contains domains with varying *c*-lattice parameter and a small fraction (<3 %) of orthorhombic phase before the transition around 85 K. The annealed p2 and p3 crystals show only sharp peaks that belong to the tetragonal phase above the transition temperatures. For p2, the structural transition occurs gradually over 40 K.

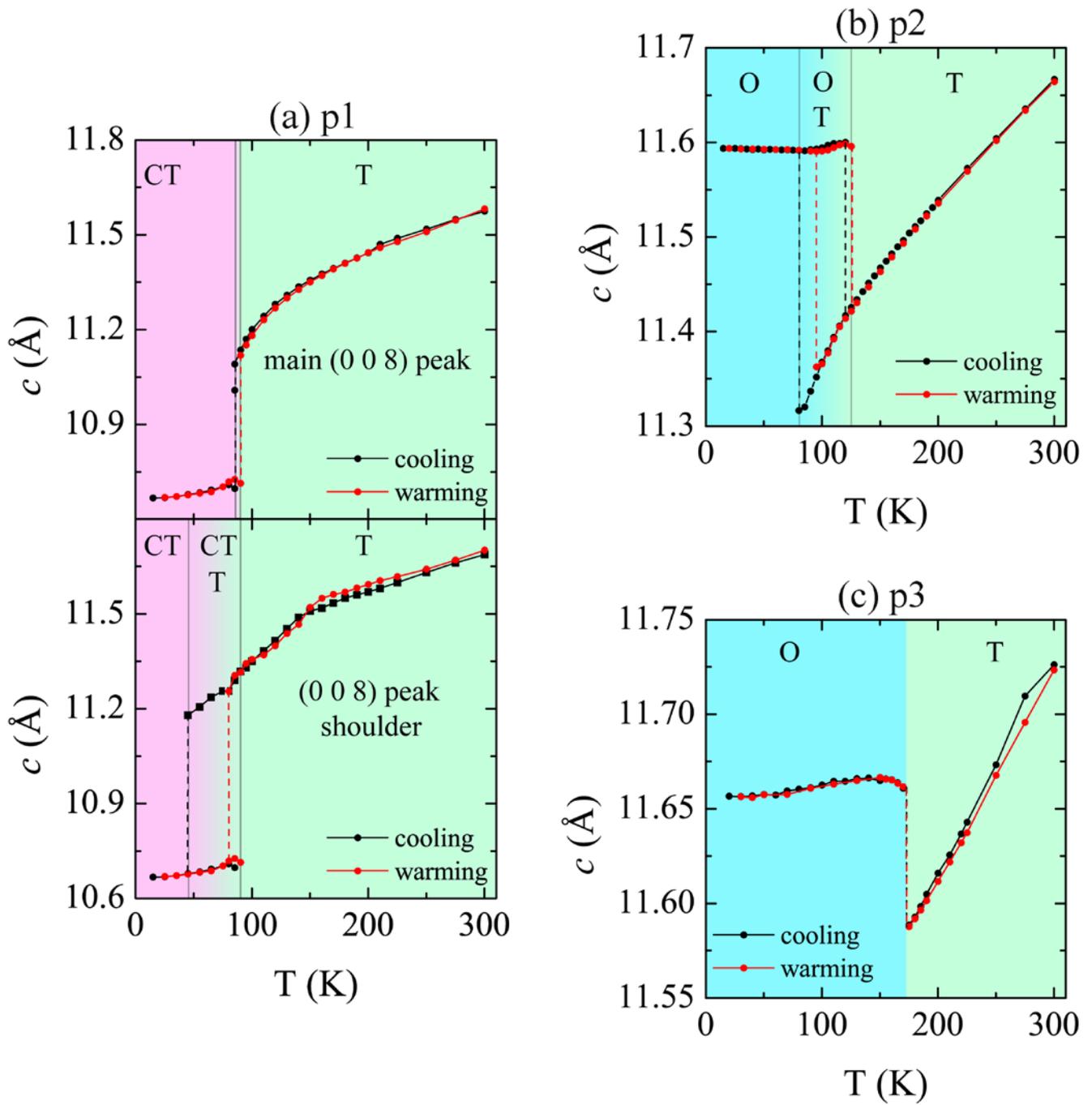

**Figure 3.** Temperature dependence of the *c*-parameter in (a) p1, (b) p2, and (c) p3 samples of $CaFe_2As_2$. High-temperature tetragonal phase (green, T) transitions to low temperature collapsed tetragonal phase (pink, CT) or orthorhombic phase (blue, O) upon cooling. See text for more details.

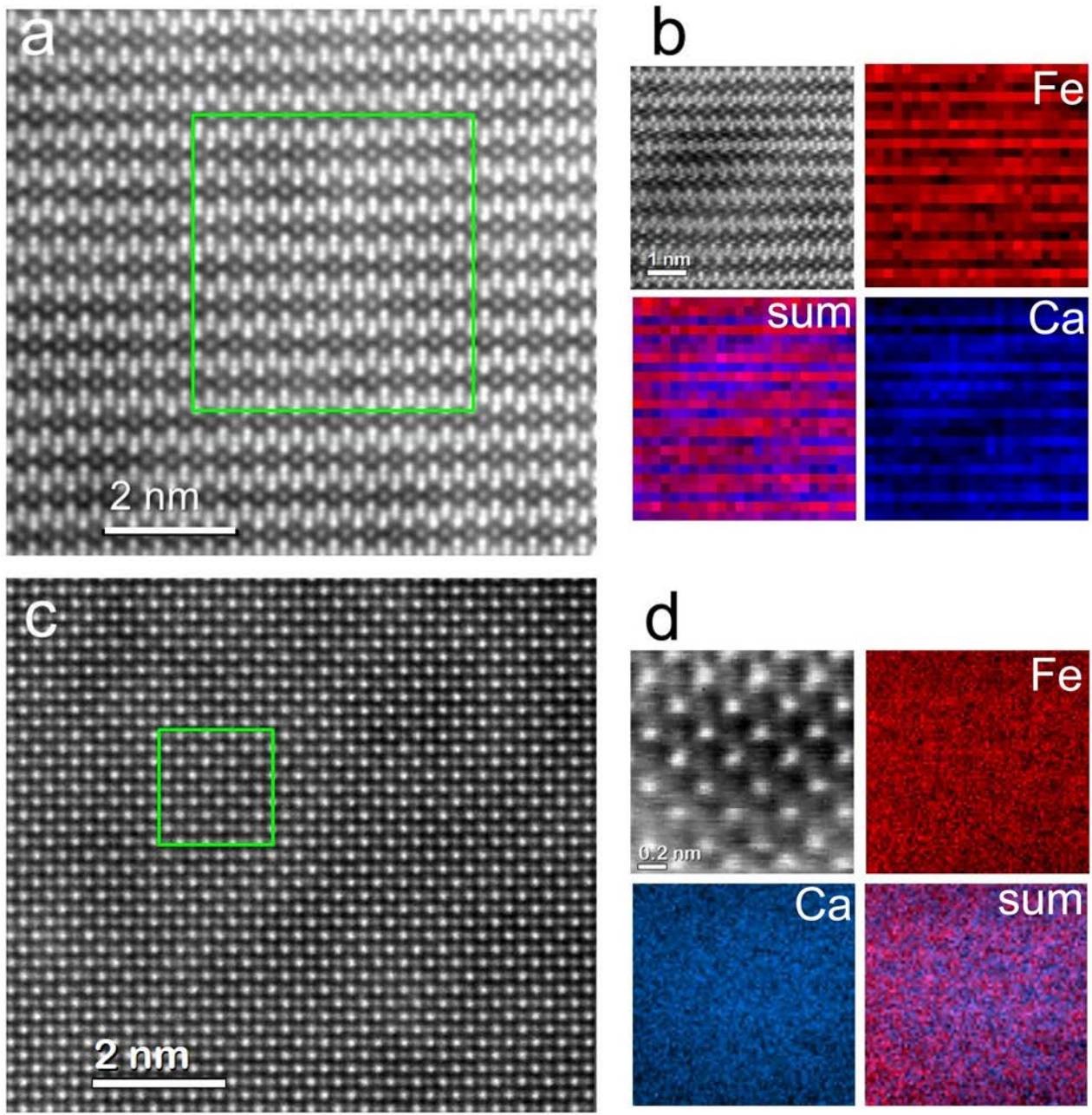

**Figure 4**. Atomic resolution HAADF images of p1 sample of $CaFe_2As_2$ with beam parallel to (a) [1 0 0] and (c) [0 0 1], and the corresponding EELS elemental maps for these regions shown in (b) and (d), respectively (shown in gray is the HAADF signal acquired simultaneously). The crystals have uniform 122 chemical compositions.

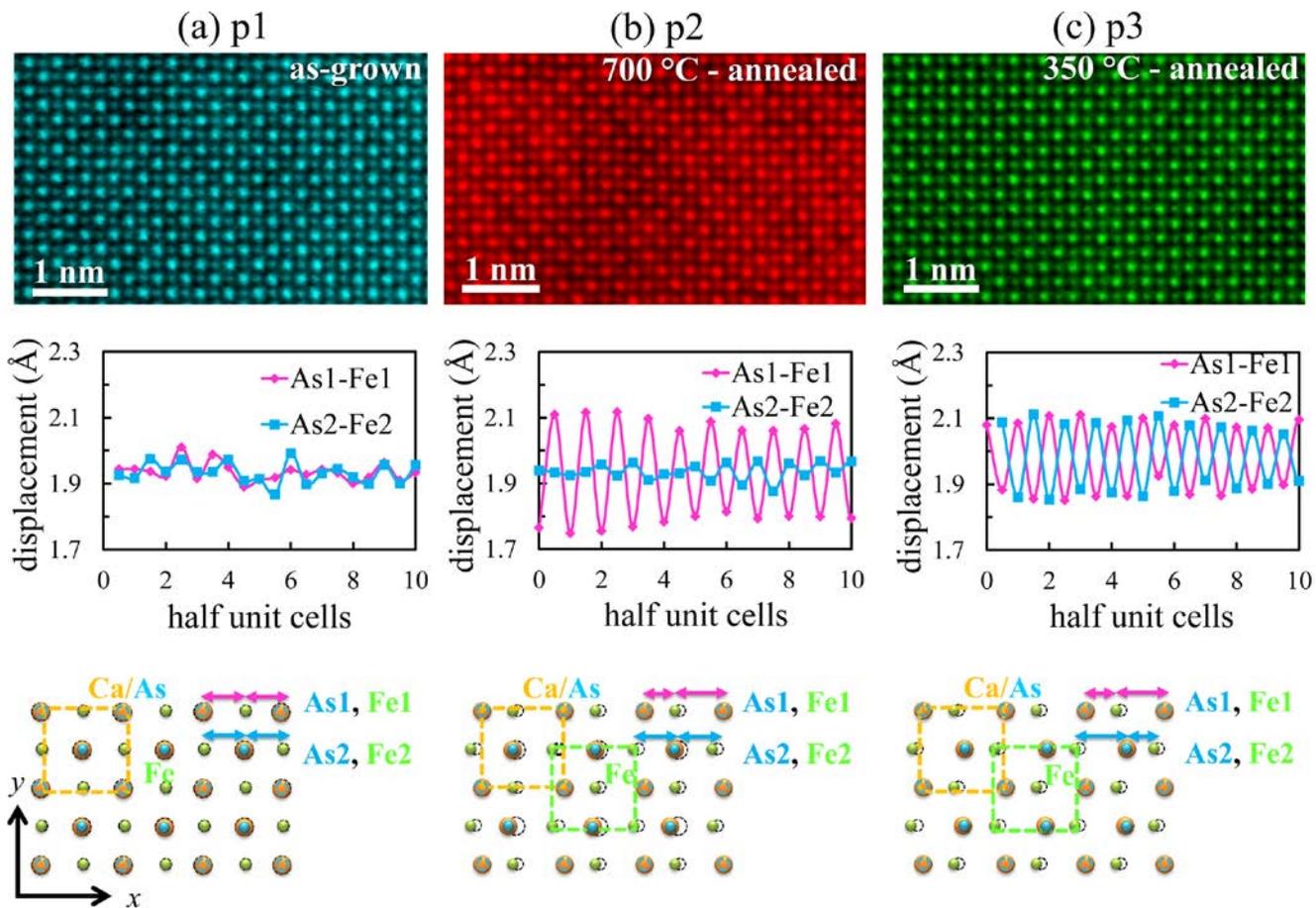

**Figure 5**. For $CaFe_2As_2$, aberration corrected HAADF images viewed normal to [0 0 1], and atomic displacements plotted as distances between As and Fe along $x$, or $|x_{As}-x_{Fe}|$ averaged in the $y$ direction in (a) p1, (b) p2, and (c) p3 crystalline samples. Strain in p1 is relieved by atomic displacements of Fe atoms upon annealing. In p3, Ca/As atoms remain in their original positions, whereas in p2, these atoms are also displaced in every second layer.

# Complex structures of different CaFe$_2$As$_2$ samples

B. Saparov*,[1], C. Cantoni[1], M. Pan[1], T. C. Hogan[2], W. Ratcliff II[3], S. D. Wilson[2], K. Fritsch[4], B. D. Gaulin[4,5,6] & A. S. Sefat[1]

[1]Oak Ridge National Laboratory, Oak Ridge, TN 37831, USA

[2]Department of Physics, Boston College, Chestnut Hill, MA 02467, USA

[3]NIST Center for Neutron Research, Gaithersburg, MD 20899-6102, USA

[4]Department of Physics and Astronomy, McMaster University, Hamilton, Ontario L8S 4M1, Canada

[5]Brockhouse Institute for Materials Research, McMaster University, Hamilton, Ontario L8S 4M1, Canada

[6]Canadian Institute for Advanced Research, 180 Dundas St W, Toronto, Ontario M5G 1Z8, Canada

*Correspondence to saparovbi@ornl.gov

## Supporting Information

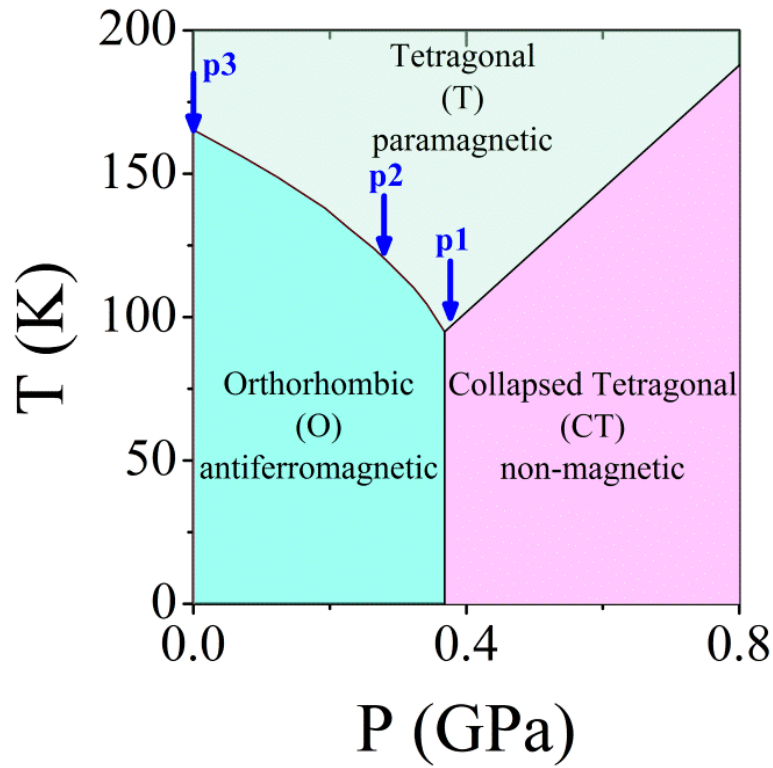

**Figure S1**. Temperature (T) – pressure (P) phase diagram of $CaFe_2As_2$; the data are adapted from references [10, 19, S1-S2]. Different parts of this phase diagram are reached here by annealing of the crystals [7]. The three samples studied in this manuscript are indicated as p1, p2 and p3. P1 undergoes a transition from paramagnetic tetragonal phase (T) to nonmagnetic collapsed tetragonal phase (CT) below 95(1) K, p2 and p3 samples transition from tetragonal to antiferromagnetic orthorhombic phases (O) at 168(1) K and 118(4) K, respectively.

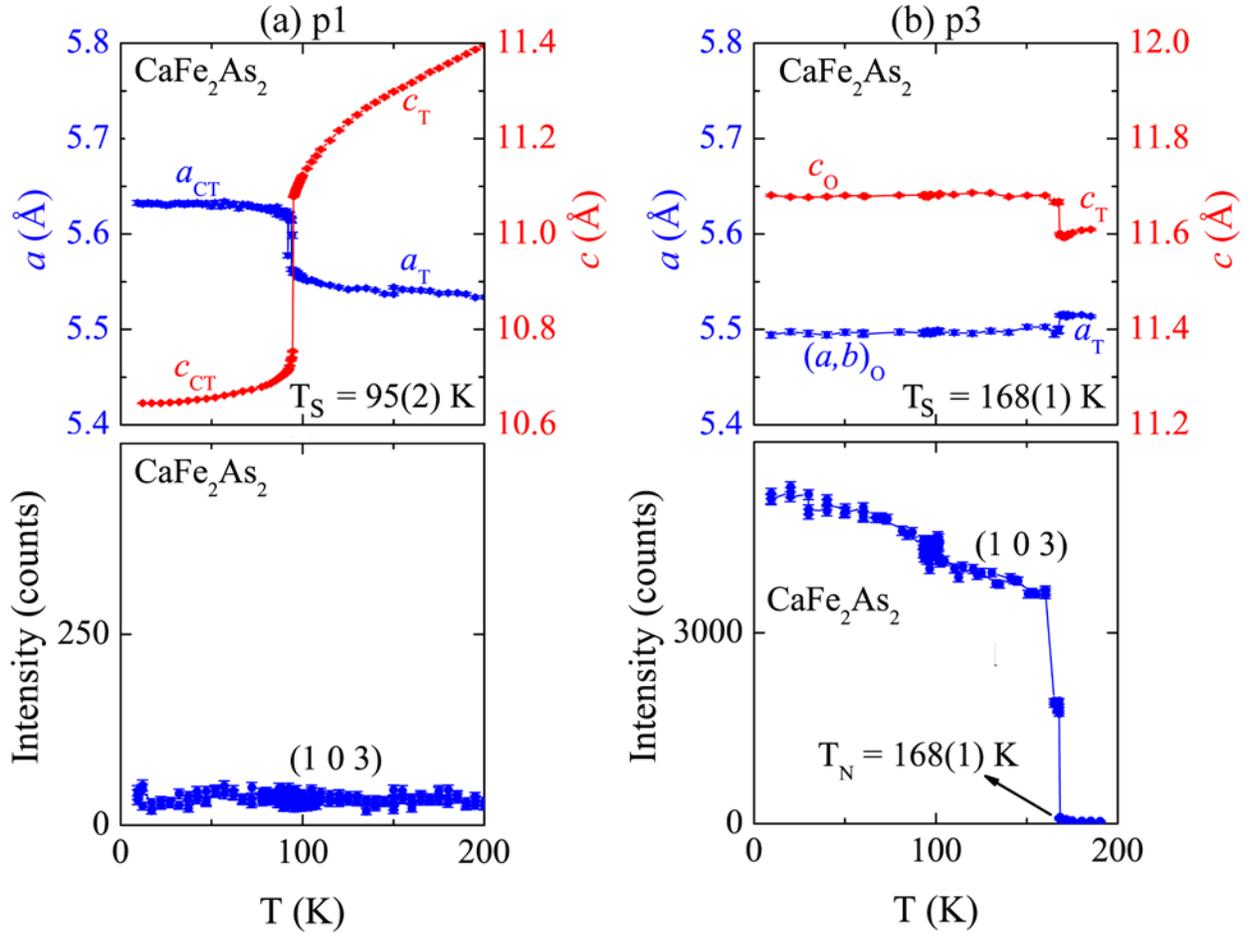

**Figure S2**. For CaFe$_2$As$_2$, temperature dependence of lattice parameters and intensity of (1 0 3) magnetic peaks obtained from neutron diffraction. (a) P1 transitions from a paramagnetic tetragonal (T) to a nonmagnetic collapsed tetragonal phase (CT) at $T_S$ = 95(2) K, whereas (b) p3 undergoes a structural transition into an antiferromagnetic orthorhombic phase (O) at $T_S = T_N$ = 168(1) K. Error bars represent +/- one standard deviation. Tetragonal $a$-lattice parameters are multiplied by √2 for comparison.

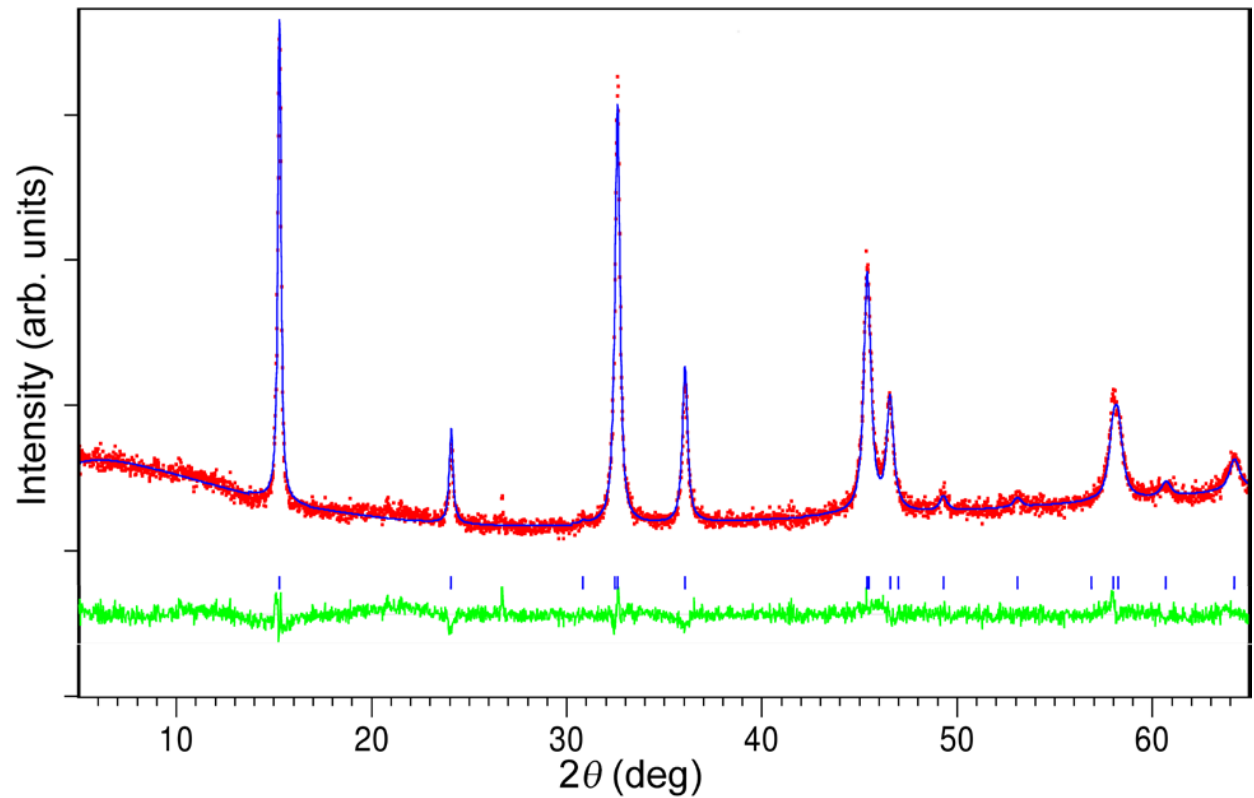

**Figure S3**. Rietveld refinement (blue line) of powder X-ray diffraction data (red dots) for as-grown p1 sample of $CaFe_2As_2$. The Bragg positions and difference plot are shown as blue tick marks and a green line, respectively.

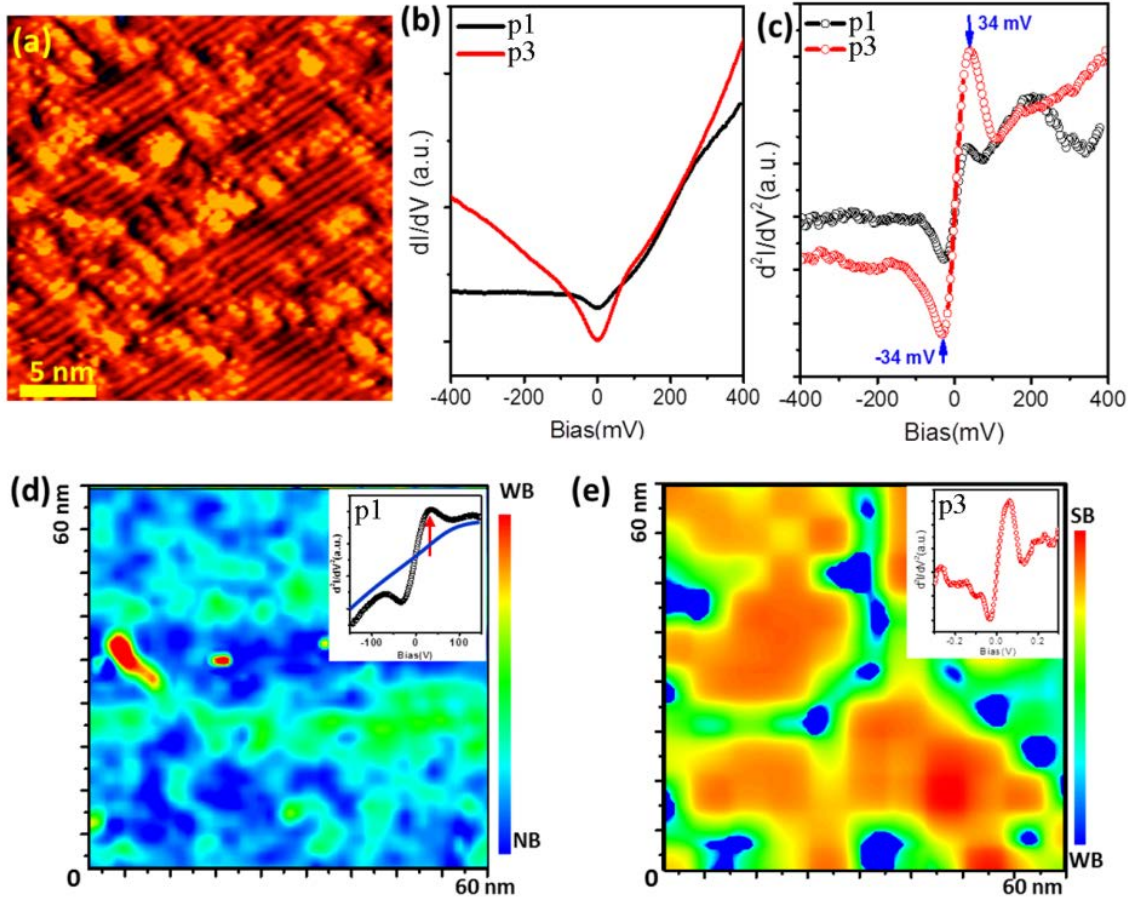

**Figure S4**. For CaFe$_2$As$_2$, (a) 24 nm × 24 nm STM topographic image for a cleaved p3 surface, taken with bias = 0.8 V, I$_t$ = 20 pA, shows "2×1" stripes. (b) Averaged dI/dV spectra observed on the surface of p1 (black curve) and p3 (red curve), showing a V-shape with small kinks located symmetrically at both positive and negative biases. Such spectra are averaged from 50 × 50 dI/dV spectra grid over a larger area of 200 nm × 200 nm size on both CaFe$_2$As$_2$ samples, in order to eliminate the effect from spatial inhomogeneity in the area. (c) The second derivative d$^2$I/dV$^2$ versus V curve, derived numerically from dI/dV spectra in the panel (b). The bosonic mode appears at 34 mV as a pair of peak and dip. Spatial mapping of bosonic mode in (d) p1 and (e) p3 samples; a slice-cut with a fixed bias at +34 mV taken from experimental d$^2$I/dV$^2$ spectra grid measurement over a larger area of 60 nm × 60 nm size. Insets of (d) and (e) show representative d$^2$I/dV$^2$ spectra for strong (SB, red), weak (WB, black), and no (NB, blue) bosonic mode regions. The spectra were taken at 78 K with bias = -100mV, tunneling current = 100pA, and modulation V =35 mV.

**Scanning Tunneling Microscopy (STM) and Scanning Tunneling Spectroscopy (STS)**: Figure S4a shows a topographic scanning tunneling microscopy (STM) image for the in situ-cleaved $CaFe_2As_2$ p3 crystal. In this image (24 nm × 24 nm), the majority of the sample surface is covered by so-called "2×1" stripes. As the in-plane lattice constant is 3.98 Å (consistent with low temperature neutron diffraction data), a high resolution image of stripes, reveals a 2×1 (8 Å × 4 Å) structure. Such 2×1 stripe structure has been frequently reported on the surface of various 122 Fe-based superconductors[S3-5], resulting from a half layer of alkaline-earth metal, after cleaving. Beside the stripes, the surface is also interrupted by some clusters, which are most likely remnants from room-temperature cleaving. To investigate the electronic difference between p1 and p3 samples, we measured the tunneling spectra at 78 K, which is well below $T_N$ = 168(1) K in p3 sample. Because point tunneling spectra by STM is a very local measurement, in order to eliminate the effect of spatial inhomogeneity, we took the differential tunneling conductance spectra (dI/dV versus V) survey on a large surface area (200 nm × 200 nm). Figure S4b shows two dI/dV spectra, for each crystal averaged over 2500 curves. Such spatial averaged spectra can represent the electronic property of the materials. Both curves exhibit a similar, asymmetric overall V-shape background. The scanning tunneling spectroscopy (STS) on p3 demonstrates prominent "kinks" located symmetrically at both positive and negative biases. The peak-dip-hump feature has been observed in STS in both cuprates[S6,S7] and FeSCs[S8,S9]. A dip-hump feature appearing at certain voltage, resulted from inelastic tunneling and is believed to reveal a bosonic excitation, probably indicating spin-fluctuation mediated superconductivity in these materials. However, it is still debated whether such bosonic mode is induced by the interaction between the electrons and a phonon mode or associated with spin excitations. Such bosonic excitation can be observed clearly in the second derivative of the tunneling spectrum, that is, $d^2I/dV^2$ versus V. Clear dips (or peaks) should be observed in $d^2I/dV^2$ curve at energies of the bosonic excitation. By taking numerical derivative from the dI/dV spectra in Figure S4b, the corresponding $d^2I/dV^2$ curves are shown in Figure S4c. Sample P3 exhibits a pair of dip/peak at the energy of ±34 meV due to strong bosonic excitations[S10] or flat optical phonon modes located at 32-34 meV[S11]. We cannot unequivocally rule out either one of these scenarios, and choose to focus on the former scenario; in this case, the peaks at ±34 meV in $d^2I/dV^2$ curve represent the integration over spin wave excitations observed in $CaFe_2As_2$[S10], which explain the broadening of inelastic spectrum. In comparison, the $d^2I/dV^2$ spectrum (black curve) on sample p1 shows a much weaker dip (or peak) feature. Contrasting the antiferromagnetic p3 and nonmagnetic p1, the different intensities of dips (or peaks) in $d^2I/dV^2$ spectra strongly imply the observed bosonic excitation is originated from spin excitations in antiferromagnetic ordering of this material, instead of phonons. Such bosonic excitation is not necessarily relevant for superconductivity, as they can be observed in the non-superconducting $CaFe_2As_2$ parent. Instead of taking the numerical derivative of the dI/dV spectra, we measured $d^2I/dV^2$ spectrum experimentally by detecting the secondary harmonic of tunneling current with a lock-in technique. In Figure S4d and S4e, the colored maps illustrate the spatial distribution of such bosonic excitation in p1 and p3 surfaces, where the value of $d^2I/dV^2$ is represented in a color scale as a function of spatial position with a fixed bias at +34 meV. It is evident that some areas show weak bosonic excitation, while other regions exhibit no-bosonic excitation. The corresponding $d^2I/dV^2$ curves are shown in the insets of Figure S4d and S4e, where the black curve shows a weak bosonic dip (or peak) and the blue curve has no bosonic feature. Therefore, we can conclude that there is a strong inhomogeneity of bosonic excitations in p1, which suggests a weak or non-antiferromagnetic ordering. These findings are consistent with our X-ray and neutron diffraction results in that there is a strongly inhomogeneous strain distribution in p1 with the minority of domains remaining in the uncollapsed magnetic state at low temperatures, while the majority of crystalline domains transition into the collapsed nonmagnetic state.